\documentclass[%
 aip,
 jmp,%
 amsmath,amssymb,
reprint,%
]{revtex4-1}

\usepackage{graphicx}
\usepackage{bm}
\usepackage[version=3]{mhchem}
\usepackage{mathrsfs} 

\usepackage{xcolor}

\makeatletter
\newcommand{\customlabel}[2]{%
\protected@write \@auxout {}{\string \newlabel {#1}{{#2}{}}}}
\makeatother

\begin{document}


\title[]{Modelling the dielectric constants of crystals using machine learning}

\author{Kazuki Morita}
\affiliation{Department of Materials, Imperial College London, London SW7 2AZ, United Kingdom }

\author{Daniel W. Davies}
\affiliation{Department of Chemistry, University College London, London WC1H 0AJ, United Kingdom}

\author{Keith T. Butler}
\email{keith.butler@stfc.ac.uk}
\affiliation{SciML, Scientific Computer Division, Rutherford Appleton Laboratory, Harwell OX11 0QX, UK}

\author{Aron Walsh}
\email{a.walsh@imperial.ac.uk}
\affiliation{Department of Materials, Imperial College London, London SW7 2AZ, United Kingdom }
\affiliation{Department of Materials Science and Engineering, Yonsei University, Seoul 03722, Korea}

\date{\today}

\begin{abstract}
  The relative permittivity of a crystal is a fundamental property that links microscopic chemical bonding to macroscopic electromagnetic response. Multiple models, including analytical, numerical and statistical descriptions, have been made to understand and predict dielectric behaviour. Analytical models are often limited to a particular type of compounds, whereas machine learning (ML) models often lack interpretability. Here, we combine supervised ML, density functional perturbation theory, and analysis based on game theory to predict and explain the physical trends in optical dielectric constants of crystals. Two ML models, support vector regression and deep neural networks, were trained on a dataset of 1,364 dielectric constants. Shapley additive explanations (SHAP) analysis of the ML models reveals that they recover correlations described by textbook Clausius--Mossotti and Penn models, which gives confidence in their ability to describe physical behavior, while providing superior predictive power.
\end{abstract}

%
%
%

\keywords{Dielectric constant, Machine learning}

\maketitle

\section{Introduction}
The dielectric response function is one of the fundamental properties of material, which can give an insight into optical and electric properties.
Specifically, the electronic high-frequency component of the dielectric constant has been explored intensively in early studies and is still of interest among researchers today.\cite{Penn1962,Shannon2006}
Numerous efforts have been made to model dielectric constants.
Most notably, the Clausius--Mossotti and Penn models, and their variants, are widely adapted throughout the literature.\cite{Shannon2006,YimK2015,Naccarato2019}

\subsection{Clausius--Mossotti Model}
The Clausius--Mossotti (CM) equation expresses the dielectric constant $\varepsilon$ as:
\begin{equation} 
  \frac{\varepsilon - 1}{\varepsilon + 2} = \sum_i \frac{4 \pi \alpha_i}{3 v}.
  \label{eqn:CM}
\end{equation}
Here, $\alpha_i$ is polarizability of atomic species $i$, and $v$ is the volume.
In the case of molecular crystals, $\alpha_i$ can be assigned to a constituent molecule,\cite{Luty1976,Miller1979,Miller1990_PNAS} and in ionic solids it is assigned to an ion.
The inherently many--body nature of the dielectric constant is reduced to this simple relation through employment of two large assumptions.

Firstly, that an external electric field is screened by a dielectric medium before reaching an atom.
In the CM model, cubic symmetry is assumed and this local field $E^{\rm loc}({\bm r})$ is expressed as
\begin{equation}
E^{\rm loc}({\bm r}) = \frac{\varepsilon + 2}{3} E^{\rm ext}({\bm r}),
\end{equation}
where $E^{\rm ext}({\bm r})$ is the external electric field.
This relation, the Lorentz relation, is frequently used for non-cubic cases and generally holds when the anisotropy is small.\cite{Tessman1953,Urano1977}

Secondly, the atomic polarizability $\alpha_i$ is assumed to be a constant additive quantity.
In other words, $\alpha_i$ is assumed to be unaffected by the surroundings.
Due to its simple form, the CM model is still being used and in practice works for many materials,\cite{Shannon2006,Korotkov2008} despite the underlying assumptions.\cite{Tessman1953,Pantelides1975,Coker1976,Fowler1985} 

Numerous efforts have been made to improve the CM model.\cite{Ruffa1963,Wilson1970,Jemmer1998}
In general, more effort was put into improving atomic polarizability $\alpha_i$ rather then changing the functional form.\cite{Ruffa1963,Wilson1970,Jemmer1998}
While Wilson and Curtis increased the accuracy of the model by considering electrostatic environment of an atom,\cite{Wilson1970} Jemmer et al. fitted parameters to M$\rm \o$ller-Plessett calculations.\cite{Jemmer1998}
Nonetheless, the range of compounds and structures in which the model is valid was found to be limited.
More recently, Shannon and Fischer have fitted atomic polarizabilities to sets of experimental and computational data, consisting of 650 values from 487 compounds.\cite{Shannon2006}
This was one of the first studies to consider compounds throughout a wider variety of structures, and their values of atomic polarizability were  more transferable.
However, due to the limited dataset, some parameters were manually fitted using chemical intuition.
Despite the success of the CM based models for many compounds,\cite{Wilson1970,Jemmer1998,Shannon2006} they are far from being universal and poorly describe covalent systems.

\subsection{Penn Model}
The premise of the Penn model is the electronic band picture.
When the overlap between electrons at different atomic sites is large, such a model is anticipated to be favorable.
Treating the external electric field as a time-dependent perturbation, Penn derived the following relation:\cite{Penn1962}
\begin{equation}
  \varepsilon \approx 1 + \left(\frac{\hbar \omega_p}{\mathscr{E}_g}\right)^2,
  \label{eqn:Penn}
\end{equation}
where $\omega_p$ is the plasma frequency and $\mathscr{E}_g$ is the width of the band gap.
The significance lies in the fact that the dielectric constant is described only by the plasma frequency and the band gap.
Using the second order perturbation theory, polarizability $\alpha$ for an isolated atom can be written $\alpha \approx (\hbar \omega_p / \mathscr{E}_g)^2$, where $\mathscr{E}_g$ is the HOMO--LUMO gap.
In this case, the screening of the external field can be ignored and Penn model is equivalent to the CM model.

As with the CM model, there are various assumptions for the Penn model.
The first main assumption is to consider the electrons as a free electron gas.
This allows the form of the equation to be greatly simplified, but is not valid when electrons are localized.\cite{Phillips1967,Phillips1968}
The second assumption is to consider only the bandgap and not the density of states around the valence and conduction band edges.
This flat band approximation results in a large error when the valence or conduction band has a large dispersion.\cite{Naccarato2019}

Various efforts have been made to improve this model.
Phillips added a correction for ionic bonds and showed that dielectric constants in zinc--blende and wurtzite were reproduced.\cite{Phillips1967,Phillips1968}
Furthermore, Van Vechten added a scaling factor taking into account that not all valence electrons contribute to the dielectric response.\cite{VanVechten1969_PR891, VanVechten1969_PR1007}
These modifications did contribute to increasing the predictive power; however, a universal model was not achieved.

\subsection{Data driven models}
The CM model approaches the problem from a molecular picture, whereas the Penn model is based on the electronic band picture of a crystal.
As such, the former has difficulty with delocalized electrons and the latter has difficulty with localized electrons. 
In both cases, the number of model parameters were increased in order to improve transferability.
In all cases, this was done empirically and the physical justification was often limited.

Rapid progress in computer technology and first-principles modelling techniques has enabled the calculation of dielectric constants for large number of materials.\cite{Butler2018_NR,dePablo2019,Dunn2020}
The modern calculation of dielectric constants are typically done using density functional perturbation theory (DFPT).\cite{YimK2015,LeeM2018,Umeda2019,Noda2020,Petousis2017,Petretto2018}
Therefore, many studies have involved analysis of datasets of dielectric constants aiming to understand trends or predict values for new materials.
For example, Han and coworkers have performed high-throughput calculation of binary and ternary inorganic compounds, and compared the relation between the band gap and the dielectric constant.\cite{YimK2015,LeeM2018}
Some studies employ machine learning (ML) methods where a statistical model is trained.\cite{Umeda2019,Noda2020}
For example, Umeda et al. trained different ML models on a dataset of 3,382 compounds.\cite{Umeda2019}
They obtain a good agreement with DFPT calculations; however, the reasons for this agreement are not discussed.

Overall, simple analytical models are limited by their  generality, whereas ML models are difficult to interpret.
By studying the difference between ML and analytical models, we may recover new underlying physics and deepen our understanding of the phenomenological behavior of the dielectric response.

In this study, we use two types of ML models and investigate the reasoning behind their predictions.
Firstly, a deep neural network and support vector regression model are trained on the same dataset.
Next, their results are compared with DFPT calculations.
Lastly, we perform Game Theoretic analysis (Shapley additive explanations) to elucidate the characteristics of the trained ML models and to investigate the reason behind their predictions.
We compare the result closely with the CM and Penn models and show that ML can indeed learn underlying physical trends, while having superior predictive power.
Additionally the Shapley analysis allows us to identify reasons for poor model performance in particular cases, which is important when considering how relaible predictions are.

\section{Methodology}
\subsection{Machine learning}
\begin{table}
  \caption{\label{tab:feature_shallow} Features used to train the support vector regression model.}
  \begin{ruledtabular}
    \begin{tabular}{cc}
      Feature & Dimensions\\
      \hline
      Band gap                             \footnote[1]{Obtained from the Materials Project \cite{Jain2013,Ong2015}} & 1\\
      $\Delta$ Pauling energy              \footnote[2]{Calculated using SMACT\cite{Davies2019_JOSS}}                & 1\\
      Material density                     \footnotemark[1] & 1\\
      Formation energy (per atom)          \footnotemark[1] & 1\\
      Oxidation state (minimum, variation) \footnotemark[1] & 2\\
      Madelung energy (minimum, maximum)   \footnote[3]{Calculated using the pymatgen package\cite{Ong2013}}    & 2\\
      Ionic species (one hot encoded)      \footnotemark[3]    & 85\\
    \end{tabular}
  \end{ruledtabular}
\end{table}

Supervised ML models were trained on a dataset of 1,364 dielectric constants.
The dataset was prepared by combining two pre-existing datasets,\cite{Petousis2017,Petretto2018} and averaging the data for the overlapping materials.
Materials with a band gap less than 0.5 eV were removed because small gap materials require very dense sampling of the Brillouin zone, which is difficult to realize in high-throughput calculations.\cite{Baroni2001}
The dielectric constant was calculated by taking a diagonal average of the electronic part of the dielectric matrix.
Since this dataset only contained minimum features, it was augmented and processed using {\it Materials Project},\cite{Jain2013,Ong2015} {\it pymatgen},\cite{Ong2013} {\it SMACT},\cite{Davies2019_JOSS} and {\it scikit-learn}.\cite{Pedregosa2011}
The dataset was split into 8:2 the training:test sets.
The same data was used when training the two different ML models.

The first ML model was support vector regression (SVR).
Before training, we analyzed the available features and removed those unimportant or similar.
This step was necessary, because having too many features increases the dimensions of the manifold that the model must learn in and thus lowers the performance.\cite{Teschendorff2018}
Specifically, we calculated feature importance using the random forest (Fig. \ref{fig:importance}) and removed low importance features: Space group (one hot encoded), atomic species (one hot encoded), number of elements, and number of sites inside the unit cell.
Next, with the criterion of $r^2 > 0.90$, we removed maximum oxidation state and variation in Maledung energy, which had high correlation with variation in oxidation state and minimum in Madelung potential, respectively (Fig. \ref{fig:corr}).
The refined features are presented in Table \ref{tab:feature_shallow}.

Four machine learning models, the random forest, gradient boost regression, kernel ridge regression, and SVR, were trained and compared (Fig. \ref{fig:all_ML} and Table \ref{tab:ml_metric_SI}).
The former two models are based on decision trees and latter two models are based on kernel methods.
Since the random forest and gradient boost regression are ensemble methods, we surmise their performance to be better, however, they were overfitting and the kernel methods were able to generalise better.
The kernel ridge regression had slightly lower performance compared to SVR, therefore SVR was employed for this study.

The second ML model was a deep neural network (DNN).
Specifically we have adopted the DNN architecture {\it MEGNet} developed by Chen, et al.\cite{ChenC2019}
{\it MEGNet} can be trained only by using the crystal structure of materials.
The exact structure of the network is presented in Fig. \ref{fig:DNN_arch}.
It is not trivial to uniquely express a crystalline system in the form of a vector, however {\it MEGNet} overcomes this difficultly by representing bonding networks as graphs and using the set2set algorithm to consistently treat different sized graphs.\cite{ChenC2019}
Information of atomic number and bond length are encoded into the graph representation.
The model was trained for 400 epochs and the layer weights of the epoch epoch with the smallest Huber loss were employed.

There are two advantages and a disadvantage of using {\it MEGNet} over SVR.
The first advantage is to avoid so called ``feature engineering''.
Feature engineering is a process of feature selection, which was necessary for SVR.
Although this can be a way of inputting our domain knowledge,\cite{Butler2018_NR} this relies heavily on experience and intuition and may obstruct systematic improvement of an ML model.
The second advantage is that it is easy to perform transfer learning.
As larger datasets are available for other materials properties, it is tempting to use that information in training.
Since the upper layers in DNNs are known to learn general trends, it is possible to import the layer weights of the upper layers in advance and improve the accuracy.\cite{ChenC2019}
Taking into account the Penn model, and the results of feature engineering for SVR (Figs. \ref{fig:corr} and \ref{fig:importance}), we performed transfer learning from {\it MEGNet} model Chen et al. trained on band gaps.\cite{ChenC2019}
The disadvantage of DNNs is that interpretability. 
Specially, because we used graph representation as the input, the relation between the features and dielectric constant were difficult to extract.

\subsection{Density functional theory calculation}
In order to validate our results, we have performed DFPT calculations on a total of 24 structures.
These structures were the materials with largest error in the SVR and DNN models.
The calculations were performed with projector-augmented wave scheme as implemented in {\it VASP}.\cite{Blochl1994, Kresse1996_PRB15, Kresse1996_PRB11169}
The structures were taken from  {\it Materials Project} and were calculated using the PBE functional.\cite{Jain2013,Ong2015,Perdew1996}
The reciprocal space were sampled so that the spacing between the $k$-point was about $2 \pi \times 0.03 \ {\rm \AA}^{-1}$.
The energy cut off was set to at least 600 eV and the wavefunctions were optimized to a tolerance of $10^{-6}$ eV.

\subsection{Shapley additive explanation}
For model interrogation, we perform Shapley additive explanations (SHAP) analysis.\cite{Lundberg2017}
The Shapley regression value is defined as:
\begin{equation}
  \phi_i =
  \sum_{S \subseteq F\backslash\{i\}} \frac
  {\left| S \right|! \left( \left| F \right| - \left| S \right| - 1 \right)!}
  {\left| F \right|!}
  \left[ f_{S_\cup\{i\}} \left( x_{S\cup\{i\}} \right) - f_S\left(x_S\right)\right] .
  \label{eqn:Shapley}
\end{equation}
Here, $S$ is a subset of the features ($F$), $i$ is a particular feature of interest, $f$ is the ML model, and $x_S$ represent values of the input features in the set $S$.\cite{Lundberg2017}
 $\phi_i$ describes how much the model output changes when feature $i$ is added to the model.
Therefore, it can be used to quantify feature importance.
As SVR cannot take features of different size, we used the median of the feature instead of removing them.
Furthermore, if we fix data $x_S$ and model $f$, we can show which features are responsible for the prediction given subset $x_S$.
Using the additivity approximation suggested by Lundberg and Lee,\cite{Lundberg2017} we calculated SHAP values of all the features and data in the dataset.

\section{Result}
\subsection{ML prediction}
\begin{figure}
  \includegraphics[width=85mm]{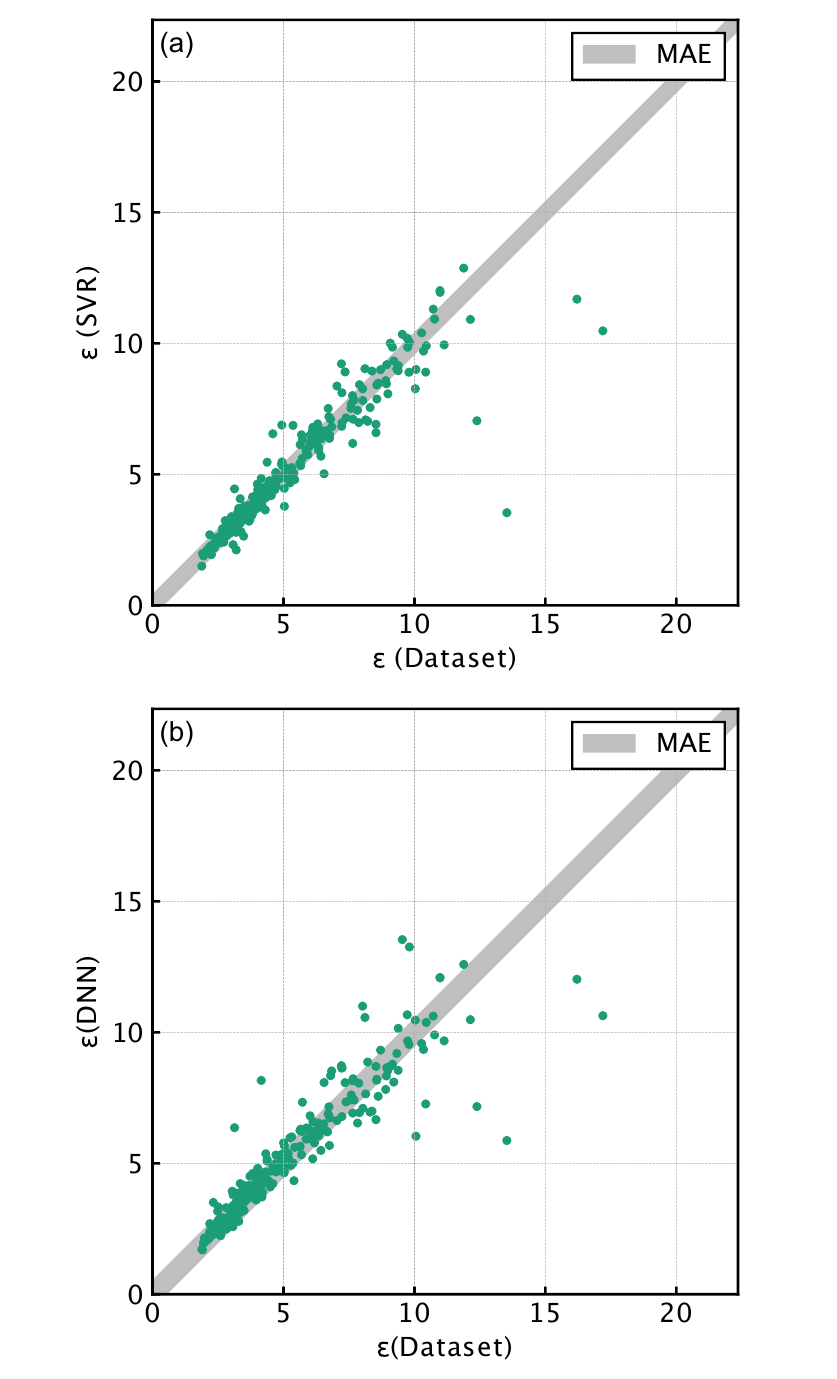}
  \caption{\label{fig:scatter} Prediction of (a) support vector machine (SVR) and (b) deep neural network (DNN) compared with dataset values. Mean average error (MAE) of the ML prediction is shaded grey.}
\end{figure}

\begin{table}
  \caption{\label{tab:ml_metric} Performance metrics of support vector regression (SVR) and deep neural network (DNN) for training and test data. Metrics are mean Pearson's correlation coefficient (r$^2$), average error (MAE), mean squared error (MSE), and root mean squared error (RMSE).}
  \begin{ruledtabular}
    \begin{tabular}{ccccc}
        & \multicolumn{2}{c}{SVR}  & \multicolumn{2}{c}{DNN}\\
      Metric & training & test & training & test \\
      \hline
      r$^2$      & 0.95 & 0.84 & 0.92 & 0.86 \\
      MAE        & 0.20 & 0.55 & 0.24 & 0.44 \\
      MSE        & 0.38 & 1.17 & 0.69 & 0.99 \\
      RMSE       & 0.62 & 1.08 & 0.83 & 0.99 \\
    \end{tabular}
  \end{ruledtabular}
\end{table}

\begin{table*}
  \caption{\label{tab:result_svr} List of 10 material with largest error in support vector regression (SVR) prediction.
  $\varepsilon_{\rm SVR}$, $\varepsilon_{\rm Dataset}$ and $\varepsilon_{\rm DFPT}$(this work) is dielectric constant from SVR prediction, training dataset and our density functional perturbation theory (DFPT) calculation, respectively.
  $\Delta \varepsilon_{\rm dataset}$ is the difference between $\varepsilon_{\rm SVR}$ and $\varepsilon_{\rm Dataset}$.
  $\Delta \varepsilon_{\rm DFPT}$ is the difference between $\varepsilon_{\rm SVR}$ and $\varepsilon_{\rm DFPT}$.
  $\Delta E_{\rm hull}$ (meV) is energy above hull per atom.\\
  }
  \begin{ruledtabular}
    \begin{tabular}{ccccccc}
      formula & $\varepsilon_{\rm SVR}$  & $\varepsilon_{\rm Dataset}$ & $\Delta \varepsilon_{\rm dataset}$ & $\varepsilon_{\rm DFPT}$ (this work) & $\Delta \varepsilon_{\rm DFPT}$ & $\Delta E_{\rm hull}$ (meV)\\
      \hline
      \ce{     LiBC} &       3.54 &     13.53 & --9.99 & 11.96 & --8.43 &  0.00 \\
      \ce{   Ga2Te5} &      10.48 &     17.19 & --6.71 & 13.75 & --3.27 &  0.00 \\
      \ce{   LiAsS2} &       7.05 &     12.39 & --5.34 &  8.57 & --1.53 &  0.00 \\
      \ce{ In2HgTe4} &      11.69 &     16.20 & --4.51 & 13.30 & --1.61 &  0.00 \\
      \ce{    KCaBi} &       9.22 &      7.21 &   2.01 &  7.51 &  1.71 &  0.00 \\
      \ce{    CdCN2} &       6.55 &      4.59 &   1.96 &  4.45 &  2.10 &  0.00 \\
      \ce{    HfNCl} &       6.88 &      4.94 &   1.95 &  4.68 &  2.21 &  0.00 \\
      \ce{    CuBS2} &       6.59 &      8.53 & --1.94 &  8.31 & --1.71 &  0.00 \\
      \ce{   CuBSe2} &       8.27 &     10.03 & --1.76 &  9.78 & --1.51 &  0.00 \\
      \ce{  Cu2HgI4} &       6.91 &      8.53 & --1.62 &  7.63 & --0.72 &  0.00 \\
    \end{tabular}
  \end{ruledtabular}
\end{table*}

\begin{table*}
  \caption{\label{tab:result_dnn} List of 10 material with largest error in deep neural network (DNN) prediction.
  $\varepsilon_{\rm DNN}$, $\varepsilon_{\rm Dataset}$ and $\varepsilon_{\rm DFPT}$(this work) is dielectric constant from DNN prediction, training dataset and our density functional perturbation theory (DFPT) calculation, respectively.
  $\Delta \varepsilon_{\rm dataset}$ is the difference between $\varepsilon_{\rm DNN}$ and $\varepsilon_{\rm Dataset}$.
  $\Delta \varepsilon_{\rm DFPT}$ is the difference between $\varepsilon_{\rm DNN}$ and $\varepsilon_{\rm DFPT}$.
  $\Delta E_{\rm hull}$ (meV) is energy above hull per atom.\\
  }
  \begin{ruledtabular}
    \begin{tabular}{ccccccc}
      formula & $\varepsilon_{\rm DNN}$  & $\varepsilon_{\rm Dataset}$ & $\Delta \varepsilon_{\rm dataset}$ & $\varepsilon_{\rm DFPT}$ (this work) & $\Delta \varepsilon_{\rm DFPT}$ & $\Delta E_{\rm hull}$ (meV)\\
      \hline
      \ce{    LiBC}  &  5.87         &  13.53 & --7.66  & 11.96 & --6.09 &   0.00 \\
      \ce{  Ga2Te5}  & 10.64         &  17.19 & --6.55  & 13.75 & --3.11 &   0.00 \\
      \ce{  LiAsS2}  &  7.17         &  12.39 & --5.22  &  8.57 & --1.41 &   0.00 \\
      \ce{In2HgTe4}  & 12.03         &  16.20 & --4.18  & 13.30 & --1.27 &   0.00 \\
      \ce{   LiZnN}  &  6.03         &  10.06 & --4.02  &  9.85 & --3.82 &   0.00 \\
      \ce{ Cs2HfI6}  &  8.17         &   4.15 &  4.01  &  3.81 &  4.35 &   0.00 \\
      \ce{    AlAs}\footnote[1]{Wurtzite structure}  & 13.54         &   9.54 &  4.00  &  9.59 &  3.95 &   6.27 \\
      \ce{    AlAs}\footnote[2]{Zinc--blende structure}  & 13.26         &   9.81 &  3.45  &  9.72 &  3.54 &   0.00 \\
      \ce{ Li3NbS4}  &  6.36         &   3.13 &  3.23  &  3.03 &  3.33 &   6.61 \\
      \ce{   MgTe2}  &  7.27         &  10.43 & --3.16  & 10.42 & --3.15 &   7.17 \\
    \end{tabular}
  \end{ruledtabular}
\end{table*}

The dielectric constants predicted by SVR are plotted against the dataset values in Fig. \ref{fig:scatter}(a).
In general, materials with dielectric constants above 10 had larger error.
DNN performed similarly to SVR as shown in Fig. \ref{fig:scatter}(b).
The main difference was suppression of error in the high dielectric constant range and slight increase of error in the low dielectric constant range in the predictions from the DNN.

Performance metrics of SVR and DNN are summarized in Table \ref{tab:ml_metric}.
This performance is similar to the performance of ML models in previous studies.\cite{Mannodi-Kanakkithodi2016,Isayev2017,Tehrani2018}
Ideally, if there is no overfitting and the ML model is trained sufficiently, metrics for training and test data should be identical, however Table \ref{tab:ml_metric} imply tendency for both models to exhibit overfitting.
For SVR, different sets of hyperparameters and loss functions were trialled and the result presented here represents the best achievable performance (Table \ref{tab:ml_metric_SI}).
We built a series of SVR models on different sized subsets of the data (Fig. \ref{fig:SVR_l_curve}) and find that the difference in the loss function metric decreased as we increased the dataset size, but did not converge. 
We therefore attribute overfitting to the small size of our dataset.
Compared to SVR, the difference in metric between the test and training data were smaller for DNN, suggesting that the latter model was able to generalize better, and exhibits overfitting to a lesser extent.
For test data, the DNN demonstrated higher performance for all of the metrics than the SVR (Table \ref{tab:ml_metric}).

The 10 materials with largest prediction error are listed in Tables \ref{tab:result_svr} and \ref{tab:result_dnn}.
Out of the 10 materials, seven of them were common amongst the two ML methods.
It is worth noting that \ce{LiBC}, \ce{Ga2Te5}, \ce{LiAsS2} and \ce{In2HgTe4} have error large enough to be identifiable in Fig. \ref{fig:scatter}.
We also present the energy above hull (obtained from {\it Materials Project}) to show that these materials are not artefacts from computational screening and could possibly be stable.\cite{Jain2013,Ong2015}

Since the dielectric constant dataset we use for training was derived from a high-throughput workflow,\cite{Petousis2017,Petretto2018} the precision is optimized to be convergent for most materials,\cite{Petousis2016} which results in some results being poorly converged.\cite{Baroni2001}
Therefore, for comparison, we have performed higher precision calculations using DFPT on these compounds and the results are presented in Table \ref{tab:result_svr} and \ref{tab:result_dnn}.

\section{Discussion}
\subsection{ML prediction}
The larger error found when the large dielectric constant is greater is not surprising (Fig. \ref{fig:scatter}).
In addition to the numerical instability in DFPT calculations,\cite{Baroni2001} there are fewer materials with large dielectric constants in the dataset, therefore the ML models were not able to fully learn the trends in the large dielectric constant range.

Comparing SVR and DNN, the performance was similar, as shown in Fig. \ref{fig:scatter} and Table \ref{tab:ml_metric}.
This is surprising since the features used for training each model were  different.
It suggests that both models are sufficiently capturing physical trends in the training data.
If the model is generalized well, we expect them to be able to detect anomalous data in the DFPT dataset.
For eight (seven) materials out of 10 materials, SVR (DNN) was actually predicts values closer to our calculations, which were performed under higher precision.
This is clearly the case for \ce{Ga2Te5} and \ce{LiAsS2}.
For example in the case of \ce{Ga2Te5} in Table \ref{tab:result_svr}, the SVR prediction was 10.48, whereas the dataset value was 17.19 and our calculation was 13.75.

Since small band gap materials require fine Brillouin zone sampling,\cite{Baroni2001} the standardized sampling in the high-throughput calculation setup may be insufficient.
To confirm this, we calculated the dielectric constant of \ce{LiAsS2} with different Brillouin zone sampling density.
Our converged value of dielectric constant was 8.57, while 11.46 was obtained using coarse sampling.
11.46 is closer to the dataset value of 12.39 (Tables \ref{tab:result_svr} and \ref{tab:k_conv}).
Although this is not a direct evidence, we speculate that materials which exhibited a large difference between our calculations and the dataset were especially sensitive to the Brillouin zone sampling and as a result, made the reported dielectric constants of these materials anomalous.

\subsection{SHAP analysis}
\begin{figure}
  \includegraphics[width=85mm]{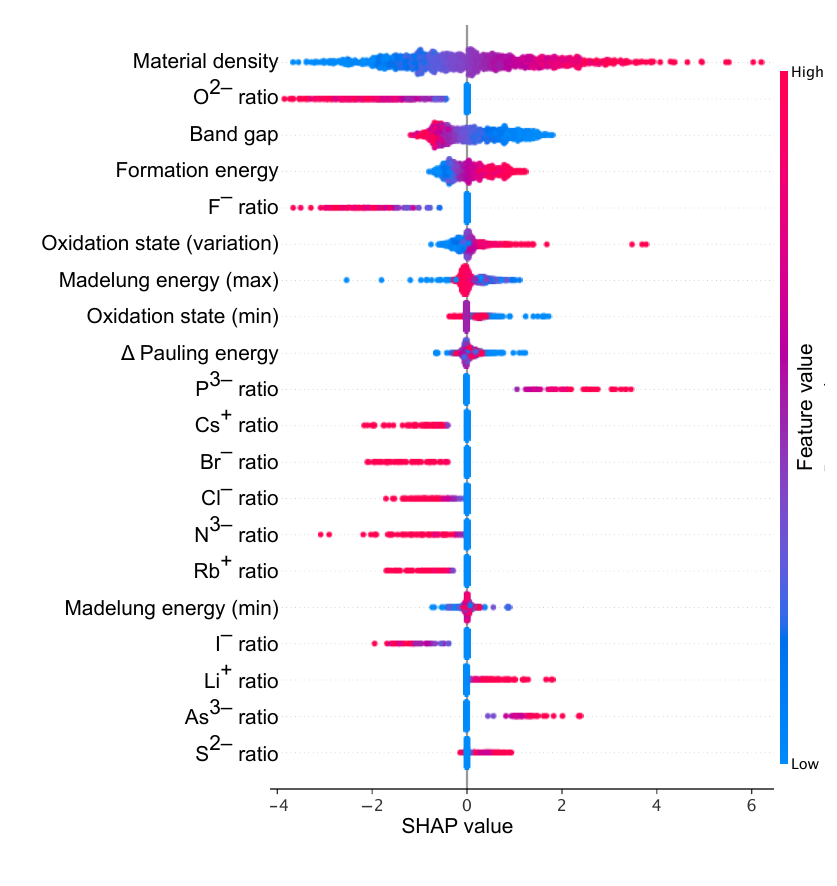}
  \caption{\label{fig:SHAP_summary} Relation of the value of each features and their SHAP values.
  Plot is colored red (blue) if the value of the feature is high (low).
  The vertical width of the plot shows number of points within the same SHAP value.}
\end{figure}

\begin{figure*}
  \includegraphics[width=170mm]{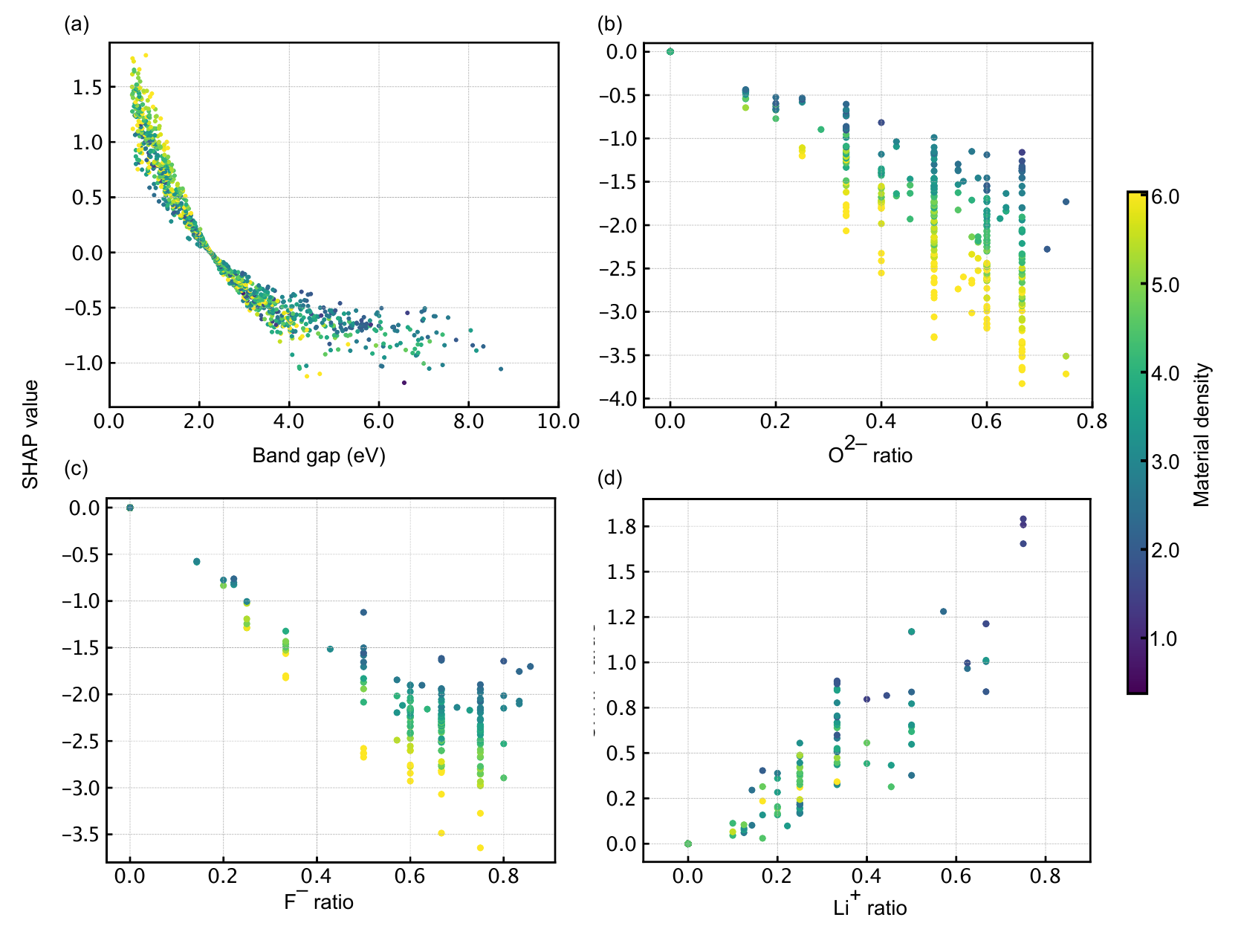}
  \caption{\label{fig:SHAP_features} SHAP values for (a) band gap, (b) \ce{O^{2-}} ratio, (c) \ce{F-} ratio, and (d) \ce{Li+} ratio. The points are colored according to material density.}
\end{figure*}

\begin{figure}
  \includegraphics[width=85mm]{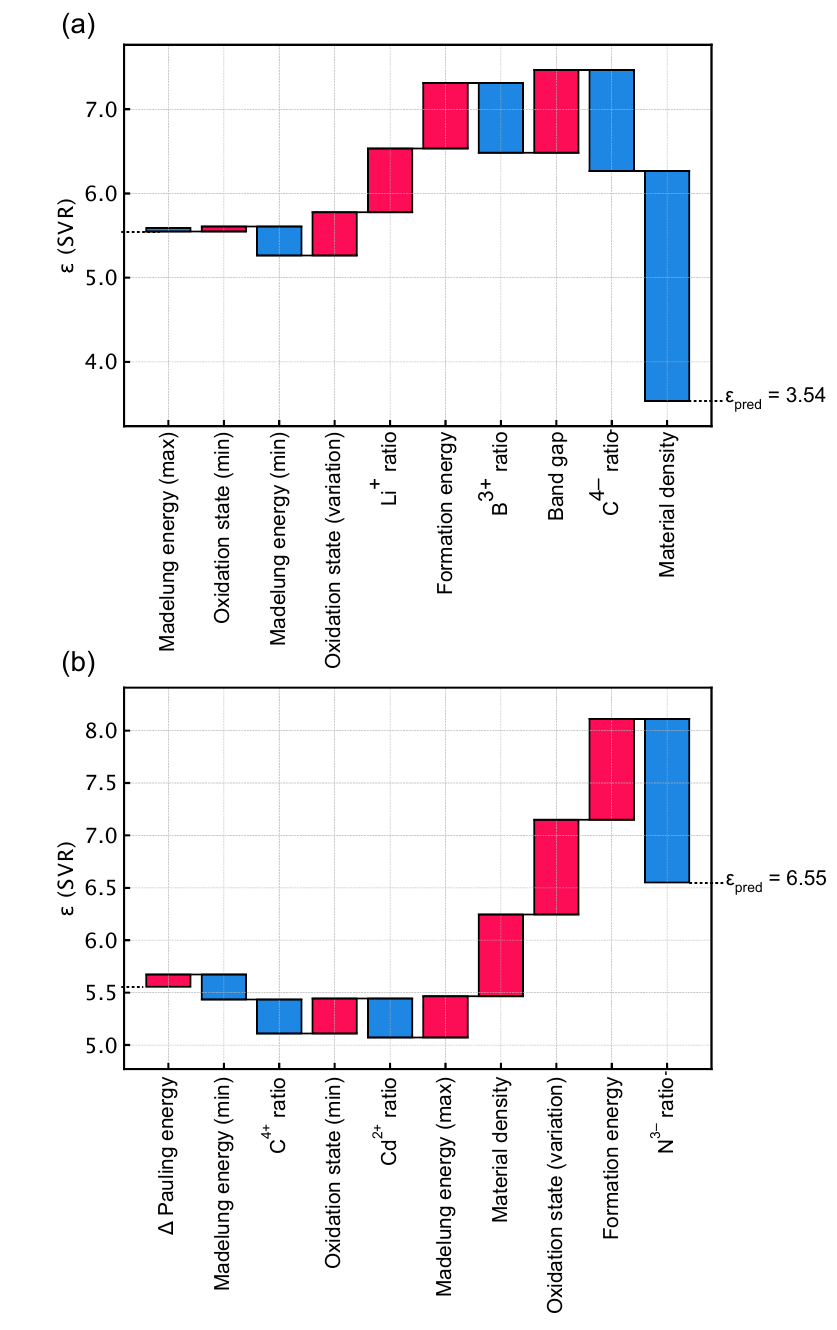}
  \caption{\label{fig:SHAP_decision_plot} 
  Breakdown of SHAP values for 10 most important features shown as a waterfall plot for (a) \ce{LiBC} and (b) \ce{CdCN2} in Table \ref{tab:result_svr}.
  Starting with the dielectric constant predicted by a median of the features ($\varepsilon$ = 5.58), each feature contributes to lowering or elevating the value of the dielectric constant in an additive manner.
  The final value of the dielectric constant $\varepsilon_{\rm pred}$ is the output of the model. 
  }
\end{figure}

SHAP analysis applies a game theoretic approach to calculate the importance of individual input features to a given model prediction. A positive (negative) SHAP value indicates that a given feature contributes to an increase (decrease) in the prediction with respect to the mean of the set.
Fig. \ref{fig:SHAP_summary} shows calculated SHAP values for all features and data.
The features are ordered by their importance.
Note that the order of importance differs from  Fig. \ref{fig:importance}, which was a random forest model trained on all features and data, whereas Fig. \ref{fig:SHAP_summary} is SVR built only on training data.

The high contribution of material density can be explained by both the CM and Penn expressions.
If there are more electrons in a given volume, the dielectric response will become larger, and indeed SHAP analysis shows that dielectric constant monotonically increases with density.

The high contribution of the band gap is also no surprise from the Penn model (Eq. \ref{eqn:Penn}).
Lower energy excitations result a larger dielectric constant. 
A large band gap gives a negative SHAP contribution.
Interestingly, the magnitude of positive contributions from small band gaps has a longer tail in the distribution than the negative contribution from large gaps.
This suggests that although a large gap decreases the dielectric response, this effect diminishes quickly so that continuously increasing band gap will not always decrease the dielectric constant. 
 
To take a deeper look at these relationships, further analysis was performed in Fig. \ref{fig:SHAP_features}(a).
The contribution from band gap decreases as the gap increases with an inverse power law relation.
By coloring the individual points according to material density, the interplay of density and band gap can be observed.
Specifically when the band gap is low, a low density increases the SHAP value, while when the gap is high a low density decreases the magnitude of the negative SHAP value.
While it is not possible to derive rigorous analytical relationship from number of data points we have, Fig. \ref{fig:SHAP_features}(a) suggests a $\phi_{\rm \mathscr{E}_g} \sim n/\mathscr{E}_g^m$ relation, where $\phi_{\rm \mathscr{E}_g}$ is the SHAP value for the band gap, $n$ is the material density, and $m$ is an arbitrary constant.
This is similar to the Penn relation (Eq. \ref{eqn:Penn}).
Therefore, it is possible to interpret that ML is learning the Penn model while incorporating other feature relations as correction terms.

When the formation energy is high (low) the dielectric constant is raised (suppressed), Fig. \ref{fig:SHAP_summary}.
This relation is opposite to that of the band gap and the negative correlation $-0.65$ between the two also agree with this trend (Fig \ref{fig:corr}).
Within our ML model, two roles of formation energy can be suggested.
Firstly, it is acting as a ionicity parameter.
According to Pauling, the formation energy of binary compounds is $\Delta H \propto (\Delta E_{\rm Pauling})^2$, where $\Delta E_{\rm Pauling}$ is the Pauling energy difference.\cite{Pauling1960}
This trend was present in our dataset (Fig. \ref{fig:SVR_formation_energy}); however, the variance was large which suggests contributions from other features.
Secondly, that it is another feature for the band gap.
In main group binary compounds, the relation $\Delta H \propto (\mathscr{E}_g^2/\mathscr{E}_f) \ln (\mathscr{E}_g/\mathscr{E}_f)$ has been reported.\cite{Phillips1970}
Quantitatively, this relation did not hold for our case (Fig. \ref{fig:SVR_formation_energy}).
Given these two relations and accounting for the fact that formation energy did not show obvious relation between the Pauling energy difference and the band gap, we suggest that it contains both information weakly.

The variation of oxidation states also exhibited high importance, Fig. \ref{fig:SHAP_summary}.
Since oxidation states have information about ionicity, we may expect it to lower the value of the dielectric constant when it is large.\cite{VanVechten1969_PR891, VanVechten1969_PR1007}
Fig. \ref{fig:SHAP_summary} shows the opposite trend.
This counter intuitive result can be explained by interpreting this feature as a correction the flat band approximation, described earlier for the Penn model.
As the band gap  is defined as a difference between the valence band maximum and the conduction band minimum, it does not have information about band dispersion.
If ionic and covalent compounds with the same band gap exist, ionic compounds will have smaller effective band gap.
When band gap is large (small) and contributes to making dielectric constant smaller (larger), lower variation in oxidation states amplifies (suppresses) this effect (Fig. \ref{fig:SVR_SHAP_oxidation_var}).
Theoretical studies also support the importance of this correction.\cite{Naccarato2019}

Other oxidation state and Madelung energy features did not show clear trends, which constitutes further evidence that the variation of oxidation state is acting as an ionicity parameter.
This result would not have been available if we were only considering feature importance, and highlights the ability of SHAP to access deeper understanding of trained models.

Since the distribution of ionic species ratio features had broad SHAP value distributions and were difficult to interpret in Fig. \ref{fig:SHAP_summary}, we continue their discussion based on Figs. \ref{fig:SHAP_features}(b--d).
SHAP values of \ce{O^{2-}} and \ce{F-} ratio are plotted against material density.
In general, higher concentration of these ionic species tend to reduce the dielectric constant.
This effect could be interpreted as originating from the strongly electronegative nature of \ce{O^{2-}} and \ce{F-}, which thus stabilize valence electrons and suppress their dielectric response.
In contrast, for the \ce{Li+} cation, the SHAP value is positive, meaning it increases the dielectric constant.
Furthermore, we can see the interplay of density and composition; when the density is higher, the effect of \ce{O^{2-}} and \ce{F-} ratio was enhanced, where \ce{O^{2-}} had slightly larger change.
Due to the limited number of data points, further study is required to draw firmer conclusions.

Finally, we explore the limitations of the trained SVR model by examining \ce{LiBC} and \ce{CdCN2} in Table \ref{tab:result_svr}.
Since SHAP values are additive,\cite{Lundberg2017} we are able to decompose them to each of the features.
We begin with the dielectric constant predicted for the median of features ($\varepsilon = 5.58$) and add contribution of all the features sequentially (Fig. \ref{fig:SHAP_decision_plot}).
Although only the 10 most important features are presented, other features had minor contribution $\sim 0.001$, so we will only discuss features presented in Fig. \ref{tab:result_svr}.
For \ce{LiBC}, which had largest error for both SVR and DNN, SVR predicted 3.54 where our calculated value was 11.96 (Table \ref{tab:result_svr}).
As shown in Fig. \ref{fig:SHAP_decision_plot}(a) the largest contribution was from the material density.
Together with the fact that the material density was small (2.15 g/cm$^2$), the error is likely due to the incorrect contribution of material density.
It is likely that in this range of material density, sufficient sampling was not achieved with our training dataset.

Most of the materials with large error feature a large contribution from material density.
The exception was \ce{CdCN2}.
SVR predicted 6.55, where 4.45 was the result of our DFPT calculation (Table \ref{tab:result_svr}).
As shown in Fig. \ref{fig:SHAP_decision_plot}(b), the three largest contributions are \ce{N^{3-}} ratio, formation energy, and variation in oxidation state, which have reasonable values.
%
We instead assign the error as arising from the large anisotropy, which is not taken into account.
The full dielectric matrix is:
\begin{equation}
  \varepsilon = \left[
    \begin{array}{ccc}
      3.40 & 0.00 & 0.00 \\
      0.00 & 3.40 & 0.00 \\
      0.00 & 0.00 & 6.55
    \end{array}
  \right]
\end{equation}
The DNN performed better because the bond lengths are explicitly taken into account through the graph representation.

\section{Conclusions}
We showed that two different machine learning models, a support vector machine and a deep neural network, were able to predict the dielectric constants of crystals with reasonable precision.
Comparison with our density functional perturbation theory calculations reveals that the machine learning models were able to detect erroneous results in the original dataset.
We performed SHAP analysis of the support vector machine model, which illustrated that it is learning similar relations to the textbook Clausius--Mossotti and Penn expressions.
Finally, we showed the limitation of our support vector model through detailed analysis of the predictions.

We suggest that as long as the dataset is sufficiently large to sample the crystal space of interest, machine learning models can be an effective approach not only for prediction of material properties but also to capture physical trends.
The analysis approach used in this study is not restricted to dielectric response, and therefore has application potential for other relations, including properties that are not as intensively studied and where there is no existing analytical description.

\section{Data Availability Statement}
A set of electronic notebooks to reproduce the model training and analysis performed in this study are available from \textit{[DOI to be inserted]}.

\begin{acknowledgments}
  The authors thank funding support from Yoshida Scholarship Foundation, Japan Student Services Organization, and Centre for Doctoral Training on Theory and Simulation of Materials at Imperial College London.
  This research was also supported by the Creative Materials Discovery Program through the National Research Foundation of Korea (NRF) funded by Ministry of Science and ICT (2018M3D1A1058536).
  Via our membership of the UK's HEC Materials Chemistry Consortium, which is funded by EPSRC (EP/L000202), this work used the ARCHER UK National Supercomputing Service (http://www.archer.ac.uk).
\end{acknowledgments}

\bibliography{dielectric_ml}

\customlabel{fig:importance}{S1}
\customlabel{fig:corr}{S2}
\customlabel{fig:DNN_arch}{S3}
\customlabel{fig:all_ML}{S4}
\customlabel{fig:CM_Penn_theory}{S5}
\customlabel{fig:SVR_l_curve}{S6}
\customlabel{fig:SVR_formation_energy}{S7}
\customlabel{fig:SVR_SHAP_oxidation_var}{S8}
\customlabel{tab:ml_metric_SI}{S1}
\customlabel{tab:hp_svr}{S2}
\customlabel{tab:k_conv}{S3}

\end{document}